\documentclass[onecolumn]{emulateapj}
\usepackage{graphicx}
\usepackage{natbib}

\renewcommand{\baselinestretch}{2}

\long\def\symbolfootnotetext[#1]#2{\begingroup%
\def\thefootnote{\fnsymbol{footnote}}\footnotetext[#1]{#2}\endgroup}

\slugcomment{}

\shorttitle{}
\shortauthors{\"Oberg et al.}

\begin{document}

\title{The effect of H$_2$O on ice photochemistry}

\author{Karin I. \"Oberg\altaffilmark{1,2}}
\affil{Harvard-Smithsonian Center for Astrophysics, MS 42, 60 Garden St, Cambridge, MA 02138}
\author{Ewine F. van Dishoeck\altaffilmark{3}}
\affil{Leiden Observatory, Leiden University, P.O. Box 9513, NL-2300 RA Leiden, The Netherlands}
\author{Harold Linnartz}
\affil{Sackler Laboratory for Astrophysics, Leiden Observatory, Leiden University, P.O. Box 9513, NL-2300 RA Leiden, The Netherlands}
\author{Stefan Andersson}
\affil{SINTEF Materials and Chemistry, P.O. Box 4760, NO-7465 Trondheim, Norway\altaffilmark{4,5,6}}

\altaffiltext{1}{Sackler Laboratory for Astrophysics, Leiden Observatory, Leiden University, P.O. Box 9513, NL-2300 RA Leiden, The Netherlands}
\altaffiltext{2}{Hubble fellow}
\altaffiltext{3}{Max-Planck Institute f\"ur Extraterrestrische Physik, Giessenbachstr. 1, D-85748 Garching, Germany}
\altaffiltext{4}{Department of Chemistry, Physical Chemistry, University of Gothenburg, SE-41296 Gothenburg, Sweden}
\altaffiltext{5}{Leiden Observatory, Leiden University, P.O. Box 9513, NL-2300 RA Leiden, The Netherlands}
\altaffiltext{6}{Gorlaeus Laboratories, Leiden Institute of Chemistry, Leiden University, P.O. Box 9502, 2300 RA Leiden, The Netherlands}

\begin{abstract}
UV irradiation of simple ices is proposed to  efficiently produce complex organic species during star- and planet-formation. Through a series of laboratory experiments, we investigate the effects of the H$_2$O concentration, the dominant ice constituent in space, on the photochemistry of more volatile species, especially CH$_4$, in ice mixtures. In the experiments, thin ($\sim$40~ML) ice mixtures, kept at 20--60~K, are irradiated under ultra-high vacuum conditions with a broad-band UV hydrogen discharge lamp. Photodestruction cross sections of volatile species (CH$_4$ and NH$_3$)  and production efficiencies of new species (C$_2$H$_6$, C$_2$H$_4$, CO, H$_2$CO, CH$_3$OH, CH$_3$CHO and CH$_3$CH$_2$OH) in water-containing ice mixtures are determined using reflection-absorption infrared spectroscopy during irradiation and during a subsequent slow warm-up. The four major effects of increasing the H$_2$O concentration are 1) an increase of the destruction efficiency of the volatile mixture constituent by up to an order of magnitude due to a reduction of back reactions following photodissociation, 2) a shift to products rich in oxygen e.g. CH$_3$OH and H$_2$CO, 3) trapping of up to a factor of five more of the formed radicals in the ice and 4) a disproportional increase in the diffusion barrier for the OH radical compared to the CH$_3$ and HCO radicals. The radical diffusion temperature dependencies are consistent with calculated H$_2$O-radical bond strengths. All the listed effects are potentially important for the production of complex organics in H$_2$O-rich icy grain mantles around protostars and should thus be taken into account when modeling ice chemistry. 
\end{abstract}


\section{Introduction}

Observations of complex molecules toward star-forming regions and in comets demonstrate the existence of efficient pre-biotic formation mechanisms \citep[e.g.][]{Belloche09,Crovisier04}.  Photochemistry in icy grain mantles
was suggested as a path to chemical complexity more than three decades ago \citep{Greenberg74}. Despite this history and a recent surge in interest, most mechanisms that produce complex molecules during UV and ion irradiation of simple ices remain poorly understood.  The photochemistry has been quantified for some pure ices \citep{Gerakines96, Oberg09d}. In astrophysical settings the chemistry is complicated by the observation that most ices are mixed. Predicting the outcome of ice irradiation in space thus requires a quantitative understanding of how different mixture constituents affect the ice photochemistry. Because of its prominence and its known effects on the ice binding environment, this study focuses on the effect of H$_2$O ice during UV irradiation of binary ice mixtures. These mixtures are not proposed to perfectly mimic the multi-component ice mixtures found in space, but are used to probe the fundamental principles of ice photochemistry; principles that can then be applied to more complicated ice systems through a combination of modeling and further experiments.

Ices are common during star formation, with H$_2$O ice reaching abundances of 10$^{-4}$ with respect to H$_2$ \citep[e.g.][]{Boogert08}. Simple ices -- H$_2$O, CO, CO$_2$, NH$_3$, CH$_4$ and CH$_3$OH -- form in molecular cloud (cores) through direct freeze-out and through hydrogenation and oxygenation of atoms -- O, C and N -- and of unsaturated molecules such as CO on grain surfaces \citep{Tielens82, Watanabe02,Ioppolo08}. From observed ice spectral features and ice maps \citep{Bergin02, Pontoppidan03, Pontoppidan06}, the formation of ices is sequential, starting with H$_2$O and CO$_2$ (and probably CH$_4$ and NH$_3$ as well). Deeper into the cloud core or later during the core contraction CO freezes out catastrophically, resulting in a second ice phase dominated by CO and later CH$_3$OH. The photochemistry of CH$_3$OH:CO ice mixtures, investigated in \citet{Oberg09d}, and of CH$_4$, NH$_3$ and CO$_2$ in H$_2$O ice mixtures are then prime targets for laboratory experiments.

Irradiation of CH$_3$OH and CH$_3$OH:CO ices result in the formation most complex C,H,O-bearing species, such as HCOOCH$_3$, CH$_3$CHO and C$_2$H$_5$OH, found in protostellar hot cores \citep{Oberg09d}. The same molecules may also form from irradiation of CH$_4$:H$_2$O ice mixtures, however, and the relative contribution from the two ice phases to the observed complex molecule abundances are unknown. Understanding the chemistry in the H$_2$O-rich ice is furthermore required to predict the formation rates of the pre-biotically important N-bearing complex molecules, since the main source of N in the ice, NH$_3$/NH$_4^+$, probably forms mixed with H$_2$O (Bottinelli et al. ApJ in press).

The response of ice mixtures to irradiation also governs the amount of e.g. CH$_4$ that remains for thermal desorption close to the protostar. Thermal desorption of CH$_4$ ice is thought to result in a complex warm-carbon-chain-chemistry around some protostars \citep{Sakai08}. One of the aims here is to investigate how the CH$_4$ ice photodestruction rate depends on H$_2$O-ice concentration and ice temperature. 

Irradiation of H$_2$O-rich ices was first investigated in the 60s and 70s, \citep[e.g.][]{Hagen79}. Since then there has been a handful of studies that have included the effects of H$_2$O on the overall photochemistry and two dedicated studies on the effect of different H$_2$O concentrations on the CH$_4$ chemistry during  proton bombardment at 10~K \citep{Moore98} and on the CH$_3$OH UV photochemistry at 3~K \citep{Krim09}. \citet{Moore98} found an increasing production of CH$_3$OH, CH$_3$CHO and CH$_3$CH$_2$OH and a decreasing production of C$_2$H$_6$ with H$_2$O concentration in H$_2$O:CH$_4$ mixtures. \citet{Krim09} found an almost constant CH$_3$OH conversion into CO, CO$_2$ and H$_2$CO for a pure and a H$_2$O:CH$_3$OH 1:1 ice, but a factor of 3-7 conversion increase in a H$_2$O:CH$_3$OH 10:1 mixture.  In both studies all changes were explained by the production of increasing amounts of OH radicals with H$_2$O concentration and subsequent radical-radical reactions or H-abstraction by OH.

H$_2$O may however affect the photochemistry in more ways than increasing the number count of OH radicals in the ice.  The binding energies in H$_2$O-rich ices are typically different compared to pure ices \citep{Collings03}, which may affect radical diffusion. Radicals may also become physically trapped in H$_2$O ice at low temperatures as is often observed for volatile molecules \citep{Collings04}. These radical-H$_2$O interactions have probably different strengths for different radicals, which may drive the chemistry in otherwise unexpected directions.

We investigate the relative importances of these potential effects of H$_2$O on the photochemistry of CH$_4$, NH$_3$ and CO$_2$ ices. The focus is on quantitative comparisons between pure and binary H$_2$O:CH$_4$ ices at different concentrations and temperatures, both during irradiation and during the subsequent warm-up. 

\section{Experiments}

The experiments are carried out on the set-up CRYOPAD \citep{Oberg05} under ultra-high vacuum conditions ($\sim$10$^{-9}$--10$^{-10}$ mbar). Ices are deposited diffusively at 18~K by introducing a gas (mixture) in the vacuum chamber along the surface normal of a gold substrate, which is temperature controlled down to 18~K with a 2~K uncertainty. 10-20 mbar gas mixtures are prepared in a separate glass manifold with a base pressure of 10$^{-4}$ mbar. The CH$_4$, NH$_3$ and $^{13}$CO$_2$ ($^{13}$CO$_2$ was used to minimize overlap with gas phase CO$_2$ features from outside of the vacuum chamber) gases have a minimum purity of 99.9\% (Indugas). Samples containing H$_2$O are prepared from the vapor pressure of de-ionized H$_2$O, further purified through freeze-thawing.

The original ice mixture as well as changes in the ice composition induced by UV irradiation are quantified through infrared spectroscopy in reflection-absorption mode (RAIRS). The relative RAIRS band strengths are consistent with relative transmission band strengths in the investigated ice thickness regime and thus certain within 20--30\% \citep{Oberg09b}. Absolute band strengths have a 50\% uncertainty, but this does not affect the quantification of the chemistry, where yields are calculated in fractions of the original ice.

The ices are irradiated by a hydrogen-discharge UV lamp, peaking at Ly-$\alpha$ but extending between 6 and 11.5 eV \citep{MunozCaro03}. All ices are irradiated with a UV flux of $\sim$1.1($\pm0.5$)$\times10^{13}$ s$^{-1}$ cm$^{-2}$ for 6 hours, resulting in a total fluence of $\sim$2.3$\times10^{17}$ cm$^{-2}$. This is comparable to the fluence an ice in a cloud core is exposed to during 10$^6$ years because of cosmic-ray induced UV photons at a flux of 10$^4$ cm$^{-2}$ s$^{-1}$ \citep{Shen04}. The lamp calibration is described in  \citep{Oberg09b}.

Table \ref{tab:n-exps} lists the photochemistry experiments in terms of their mixture composition, ice temperature during UV irradiation and total thickness.  Following irradiation at 20--100~K, all experiments are heated by 1 K min$^{-1}$ to 150--250~K, while acquiring RAIRS every 10 min. The ice thicknesses range between $\sim$15--54 monolayers (ML), but most experiments are carried out with $\sim$40~ML ices. This ice thicknesses are similar to what is expected in the dense and cold stages of star formation and they are also in the linear regime for RAIRS. The ices are also thick enough that photodesorption will not significantly affect the bulk of the ice, based on previously measured UV photodesorption yields \citep{Oberg09b, Oberg09c}.

\section{Results}

The following five sub-sections focus on 1) the photodestruction cross sections of CH$_4$, NH$_3$ and CO$_2$, 2) the identification of the photoproducts, 3) the H$_2$O:CH$_4$ photochemistry as a function of H$_2$O concentration during irradiation and (4) during warm-up, and 5) the effects of temperature on the H$_2$O:CH$_4$ UV photochemistry in dilute mixtures.

\subsection{Photodestruction cross sections in pure and mixed ices}

\begin{figure}
\plotone{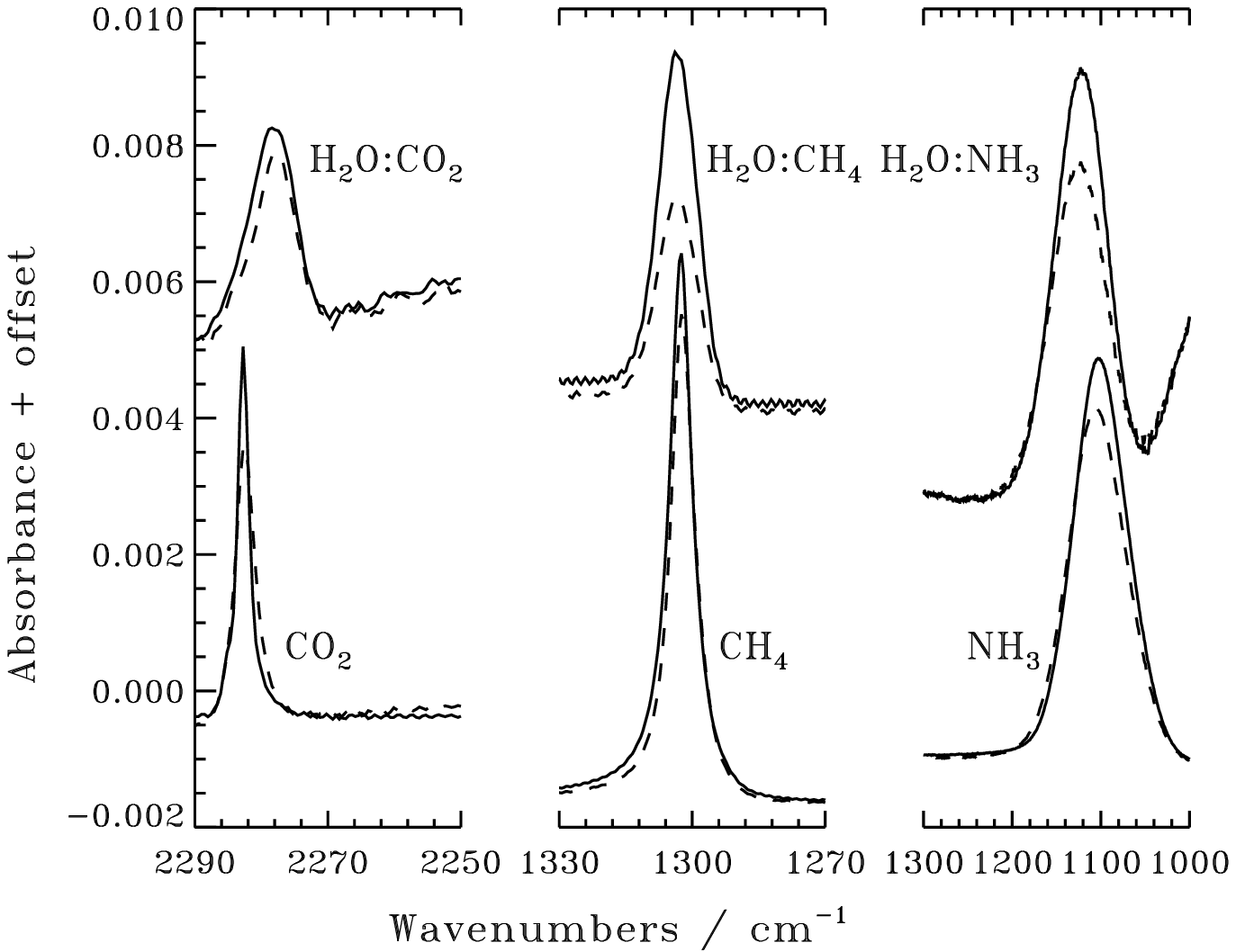}
\caption{A $^{13}$CO$_2$, a CH$_4$ and a NH$_3$ ice spectral feature before (solid) and after (dashed) a UV fluence of $2.3\times10^{17}$ cm$^{-2}$ at 20--30~K for pure ices (bottom panel) and for H$_2$O:X $\sim$ 5:1 ice mixtures (top panel). \label{fig:dis_spec}}
\end{figure}

The photodestruction rate of a species determines the production rate of radicals in the ice. Its maximum value is the photodissociation rate, as measured in the gas-phase. Previous ice experiments have however shown that the measured ice photodestruction rates are substantially lower than the gas-phase photodissociation rates \citep{Cottin03}, probably because of fast back-reactions between the dissociation fragments. Figure \ref{fig:dis_spec} shows the $^{13}$CO$_2$, CH$_4$ and NH$_3$ spectral features before and after a UV fluence of $2.3\times10^{17}$ cm$^{-2}$ in pure ices and H$_2$O:X $\sim$5:1 ice mixtures at 20--30~K. During irradiation, more CH$_4$ and NH$_3$ are destroyed in the mixtures than in the pure ices.

\begin{figure}
\plotone{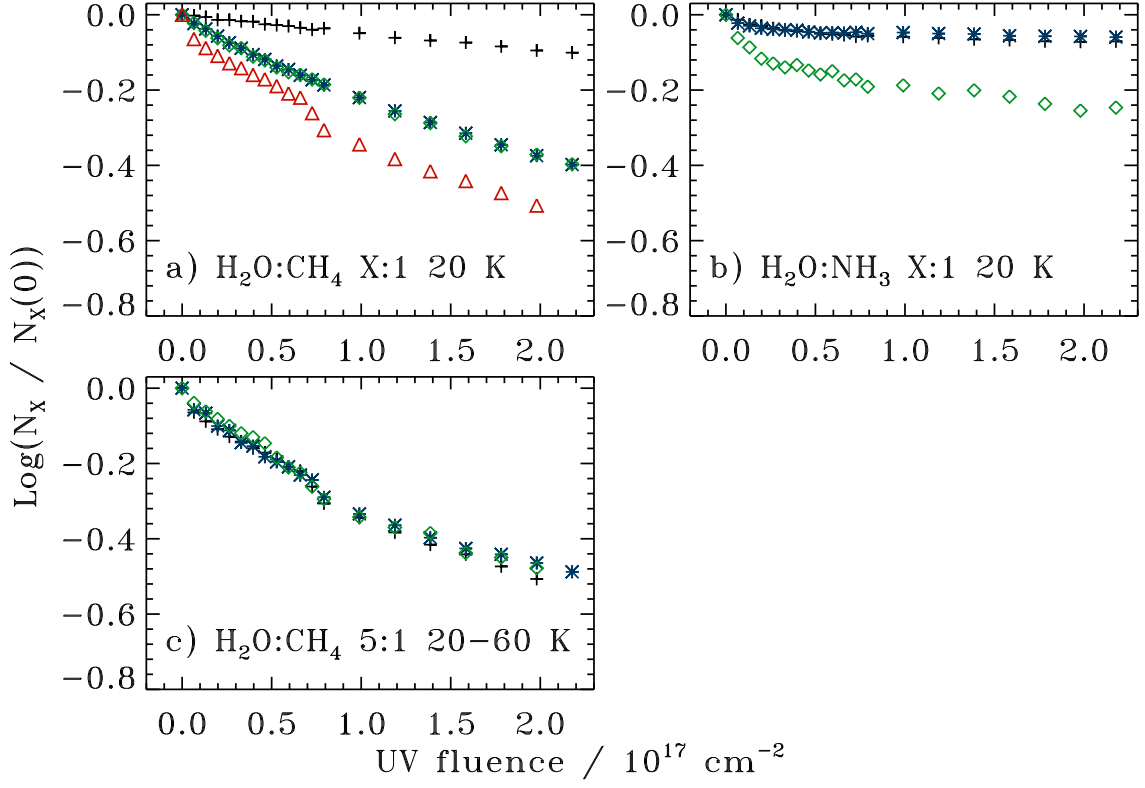}
\caption{The normalized and log-transformed loss of CH$_4$ and NH$_3$ in a) H$_2$O:CH$_4$ ice mixtures at 20-30~K (crosses = pure ice, stars = 1:3, diamonds = 2:1, triangles = 4:1), b) H$_2$O:NH$_3$ ice mixtures at 20-30~K (crosses = pure ice, stars = 2:1, diamonds = 5:1) and c) H$_2$O:CH$_4$ 5:1 ice mixtures at 20-60~K (crosses = 20~K, stars = 40~K, diamonds = 60~K). \label{fig:dis_quant}}
\end{figure}

This effect is shown quantitatively in Fig.  \ref{fig:dis_quant}, where the loss of CH$_4$ and NH$_3$  in the different ices is plotted as a function of UV fluence (overlap with gas-phase CO$_2$ lines prevents a similar analysis for CO$_2$ ice). The photodestruction cross sections are calculated from the first $4\times10^{16}$ photons cm$^{-2}$, where the curves are still approximately linear, and the results are listed in Table \ref{tab:n-exps} -- the listed uncertainties are absolute, the fit uncertainties are 10-20\%. The photodestruction cross sections for pure CH$_4$ and NH$_3$ are 0.5 and 1.4$\times10^{-18}$ cm$^2$, respectively. Both are an order of magnitude lower than the measured gas phase photodissociation cross sections \citep{vanDishoeck88}. They are still higher than previously reported values for 1000~ML ices \citep{Cottin03}, suggesting that those experiments may have suffered from optical depth effects. 

The photodestruction cross sections are up to an order of magnitude higher in the H$_2$O ice mixtures, close to the gas phase photodissociation values. In the case of CH$_4$, even the small amount of H$_2$O in the H$_2$O:CH$_4$ 1:3 ice mixture results in a factor of five higher destruction cross section, while a H$_2$O-dominated ice (5:1) is required to increase the NH$_3$ cross section significantly.

The increasing UV photodestruction with H$_2$O concentration suggest that H$_2$O molecules surrounding the photodissociated volatiles may trap some of the radicals and thus inhibit reformation of CH$_4$ and NH$_3$. The reason behind the differences between different species is difficult to assess without further modeling, but the effect seems larger, the lighter the volatile dissociation radicals (CH$_3$+H vs. NH$_2$+H vs. CO+O), assuming that the lack of an effect of H$_2$O on the final CO$_2$ dissociation fraction corresponds to a constant CO$_2$ photodissociation cross section with H$_2$O concentration. 

Temperature does not affect the CH$_4$ photodestruction cross section between 20 and 60~K for dilute ($\sim$5:1) H$_2$O:CH$_4$ ice mixtures (Fig. \ref{fig:dis_quant}c). The effect on more CH$_4$-rich ice mixtures cannot be tested because of thermal outgassing of CH$_4$ above 30~K when in high concentrations. In comparison, recent experiments on pure CH$_3$OH ice photodissociation resulted in an increasing photodestruction (then termed `effective photodissociation') cross section with temperature by up to 50\% \citep{Oberg09d}. In the CH$_3$OH study this was explained by an increased escape probability of the dissociation fragments at higher temperatures, thus lowering the back-reaction rate. A similar effect was expected for CH$_4$. Its absence suggests that thermal diffusion is fast already at 20~K for the volatile CH$_4$ dissociation fragments compared to the main CH$_3$OH photodissociation product, CH$_2$OH. 

\subsection{Product identification}

\begin{figure}
\plotone{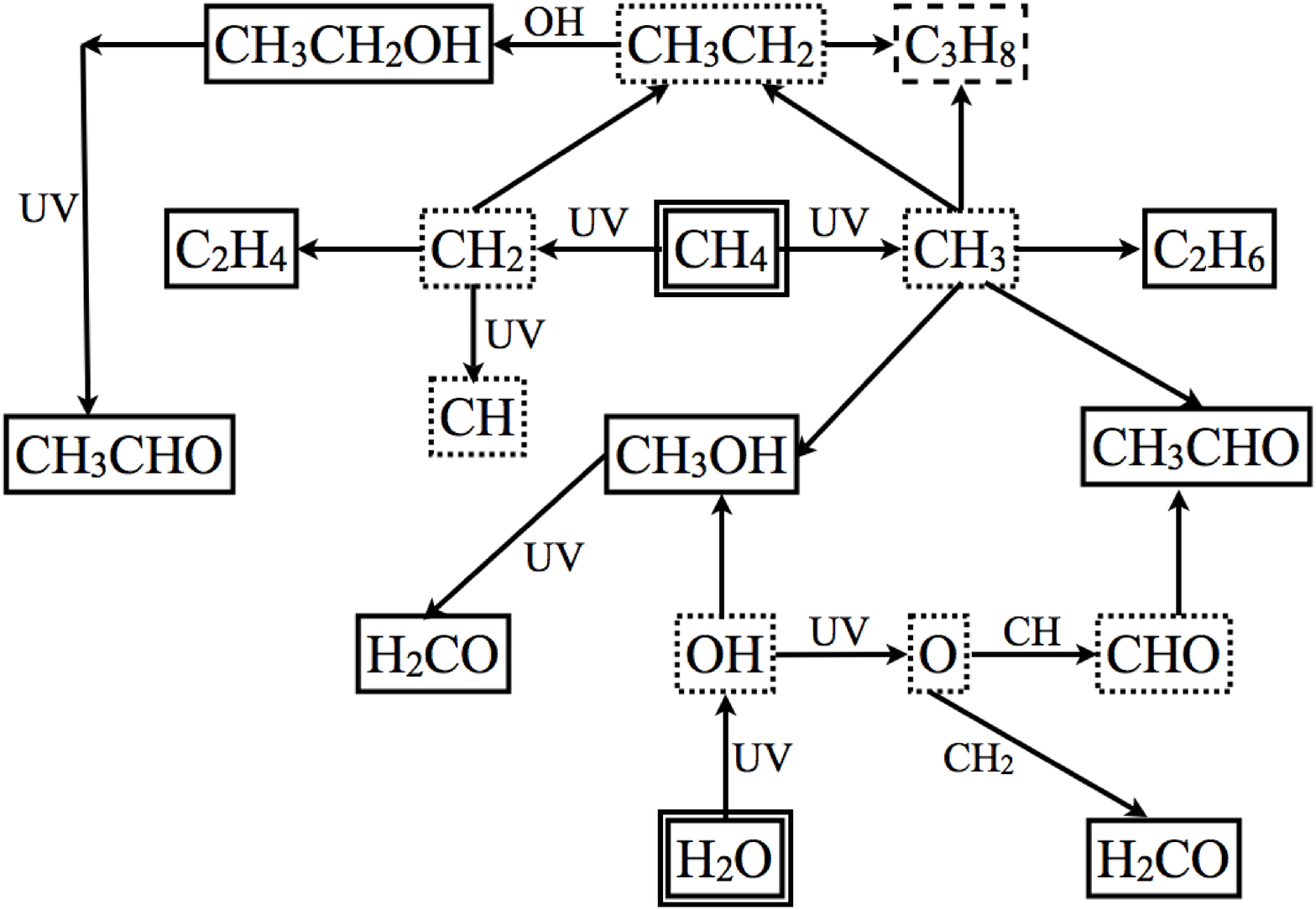}
\caption{Reaction scheme of expected main reaction pathways during UV irradiation of H$_2$O:CH$_4$ ice mixtures. In addition, small abundances of the CH$_3$OH photoproducts shown in \citet{Oberg09d} may form. Solid boxes mark species securely identified in the experiments. \label{fig:rs}}
\end{figure}

\begin{figure}
\plotone{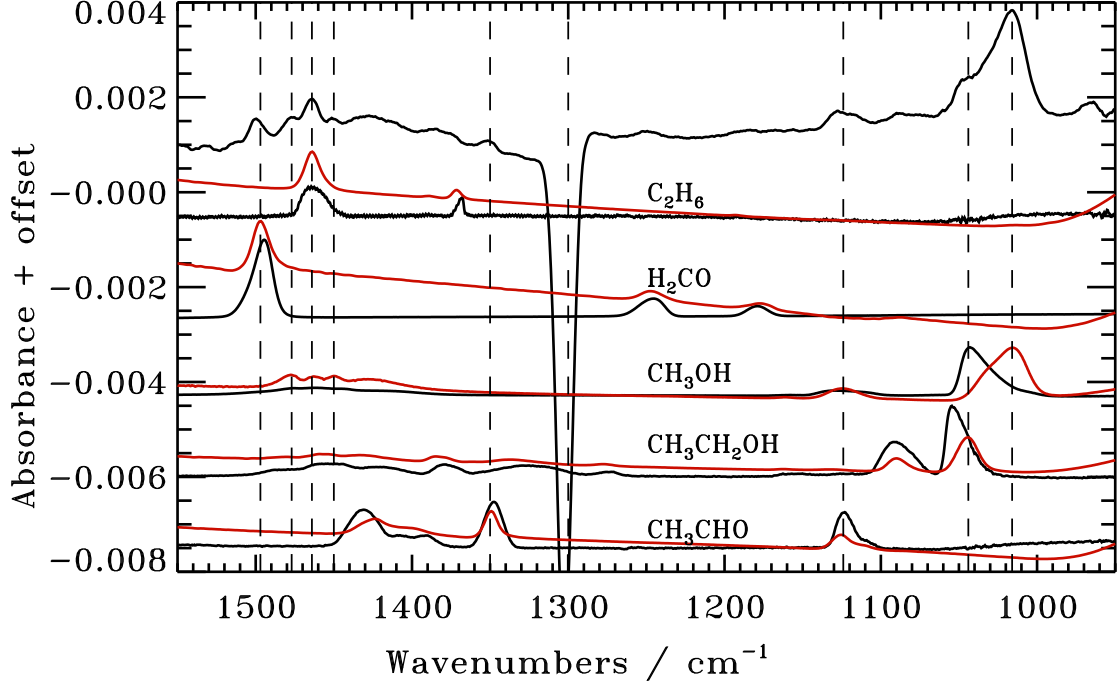}
\caption{The spectrum of an irradiated H$_2$O:CH$_4$ 2:1 mixture at 20~K is plotted on top together with pure (black) and H$_2$O mixtures (red) of expected products. The H$_2$O mixture spectra are taken from the NASA Goddard Cosmic ice laboratory spectral database by Moore et al. (http://www-691.gsfc.nasa.gov/cosmic.ice.lab). \label{fig:id_spec}}
\end{figure}

Quantification of the ice photochemistry requires secure identifications of the main reaction products. This was not possible for the UV-irradiated H$_2$O:NH$_3$ and H$_2$O:CO$_2$ ices, because the spectral features of expected main products -- NH$_2$OH, HCOOH, H$_2$CO$_3$ -- are blended with bands of H$_2$O, NH$_3$ or other photoproducts. The expected photoproducts in H$_2$O:CH$_4$ ices are shown in Fig. \ref{fig:rs}. Figure \ref{fig:id_spec} shows that most of the spectral features appearing upon irradiation of H$_2$O:CH$_4$ ice mixtures can be assigned to the predicted main products: C$_2$H$_6$, H$_2$CO, CO (not shown), CH$_3$OH, CH$_3$CH$_2$OH and CH$_3$CHO, in agreement with \citet{Moore98}. A few bands remain unassigned and are probably due to more complex molecules and to radicals. The figure also shows that two of the products, CH$_3$OH and CH$_3$CH$_2$OH, have bands that shift significantly between the pure ice and the H$_2$O-mixture. This introduces some uncertainty in the assignment of CH$_3$CH$_2$OH -- it is only possible to securely separate the relative contributions of CH$_3$OH and CH$_3$CH$_2$OH in a subset of the experiments.

In general the spectral bands used for identification and quantitative analysis have minimum overlap with spectral features of other detected species as well as of more complex molecules \citep{Oberg09d}. These band positions and  strengths are listed in Table \ref{tab:bands}.  When deriving ice abundances, the region of the investigated bands is typically fitted with several Gaussians to account for band overlaps. This approach was chosen instead of fitting ice spectra because of the varying spectral shapes in different ice environments.  The same band strengths were adopted for all ice mixtures, since the exact band strengths are unknown for most of the used ice mixtures. 

\subsection{H$_2$O concentration effects at 20~K}

\begin{figure}
\plotone{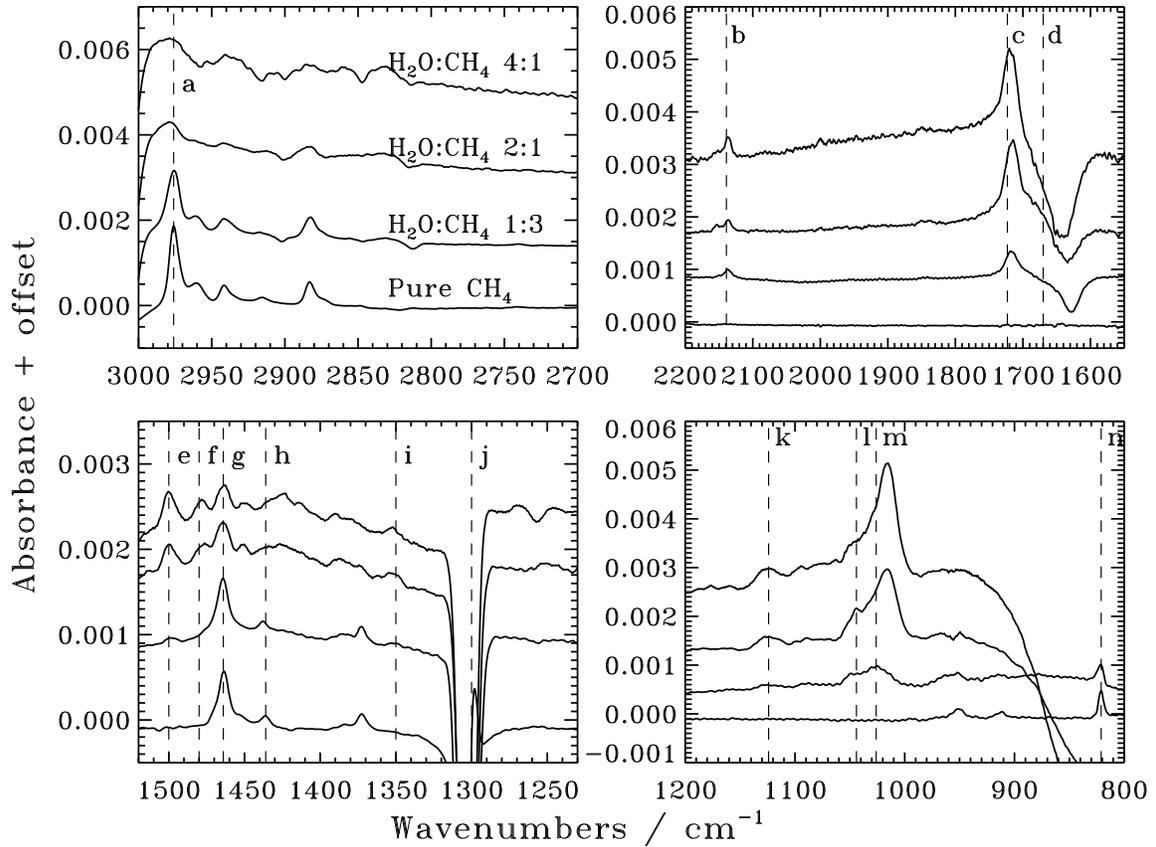}
\caption{Difference spectra of CH$_4$ ice and H$_2$O:CH$_4$ ice mixtures after a UV fluence of $2.3\times10^{17}$ photons cm$^{-2}$. The marked peaks are a) C$_2$H$_6$, b) CO, c) XCHO, d) H$_2$O, e) H$_2$CO, f) CH$_3$OH, g) C$_2$H$_6$, h) C$_2$H$_4$, i) CH$_3$CHO, j) CH$_4$, k CH$_3$OH, l) CH$_3$CH$_2$OH, m) CH$_3$OH and n) C$_2$H$_4$. \label{fig:conc_spec}}
\end{figure}

Figure \ref{fig:conc_spec} shows that irradiation of pure CH$_4$ ice results in the production of C$_2$H$_6$, C$_2$H$_4$ and larger hydrocarbons, as expected -- no C$_2$H$_2$ was observed and in general the products are hydrogen-rich. Adding H$_2$O to the CH$_4$ ice results in the appearance of CO, H$_2$CO, CH$_3$OH, CH$_3$CHO and CH$_3$CH$_2$OH spectral features. The same bands are visible in all H$_2$O ice mixtures, but the relative contributions of different bands change with concentration.

\begin{figure}
\plotone{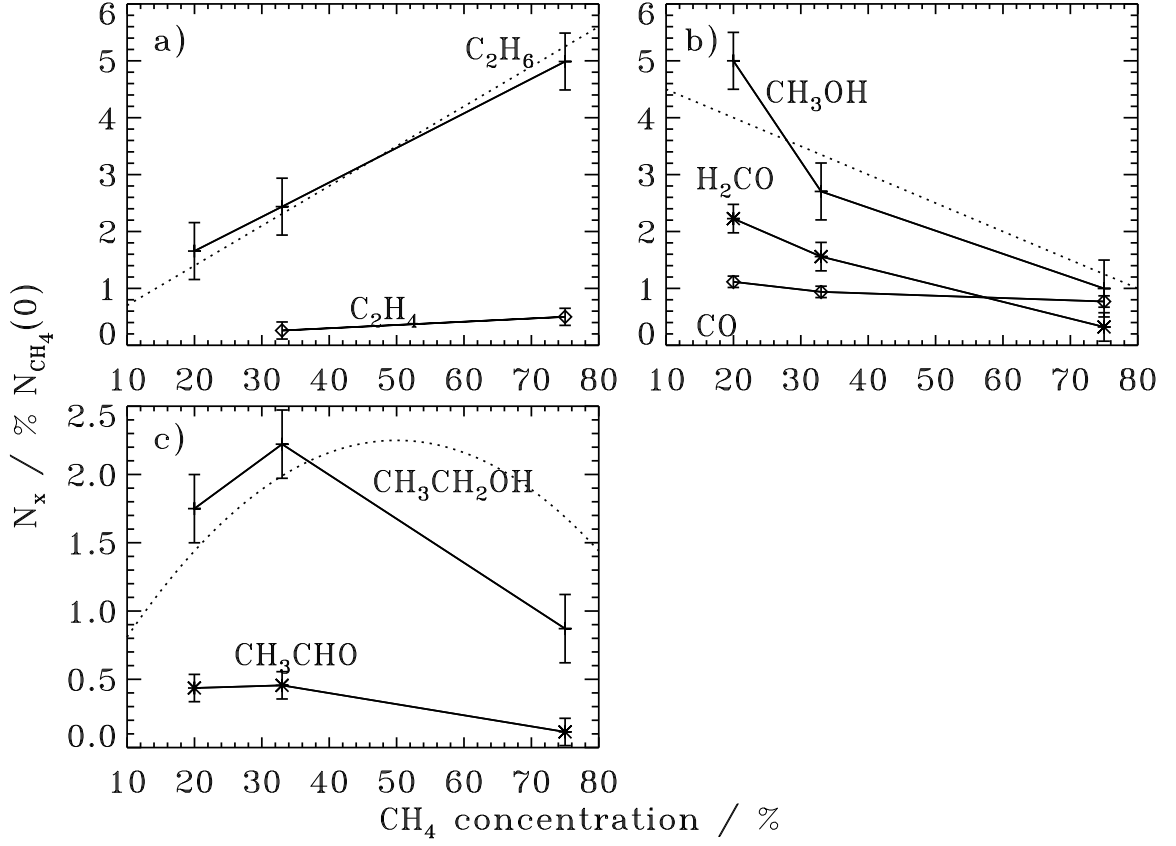}
\caption{The normalized, final abundances of the identified photoproducts as a function of CH$_4$ concentration in the ice. The error bars only include the relative uncertainties, the absolute uncertainties are about a factor of two. The dotted lines are the qualitative predictions using Eqs. (1)--(3). \label{fig:conc_fin}}
\end{figure}

The effect of H$_2$O concentration on the final product abundances, with respect to CH$_4$, is quantified in Fig. \ref{fig:conc_fin}. The identified products can be sorted into three groups a) hydrocarbons which have increasing abundances with CH$_4$ concentration, b) small organics like CH$_3$OH that form from a CH$_4$ and a H$_2$O fragment, and c) larger organics like CH$_3$CH$_2$OH that form from two CH$_4$ fragments and one H$_2$O fragment. The molecules in each group are then expected to form as:

\begin{eqnarray}
\frac{N_{\rm C_2H_6}}{N_{\rm CH_4}}\propto {\rm[CH_4]},\\
\frac{N_{\rm CH_3OH}}{N_{\rm CH_4}}\propto (100-{\rm[CH_4]}) \: {\rm and}\\
\frac{N_{\rm CH_3CH_2OH}}{N_{\rm CH_4}}\propto (100-{\rm[CH_4]}){\rm[CH_4]}.
\end{eqnarray}

Qualitatively the abundance patterns of these three molecules agree with the curve predictions. The difference in e.g. the peak position of the curve in Fig. \ref{fig:conc_fin}c indicates that the effects of diffusion and different photodestruction cross sections for different mixtures cannot be neglected when making quantified predictions, however. In addition, the production dependencies of H$_2$CO, CO and CH$_3$CHO are expected to be complicated by multiple formation pathways, all involving multiple dissociation events as shown in Fig. \ref{fig:rs}.

\begin{figure}
\plotone{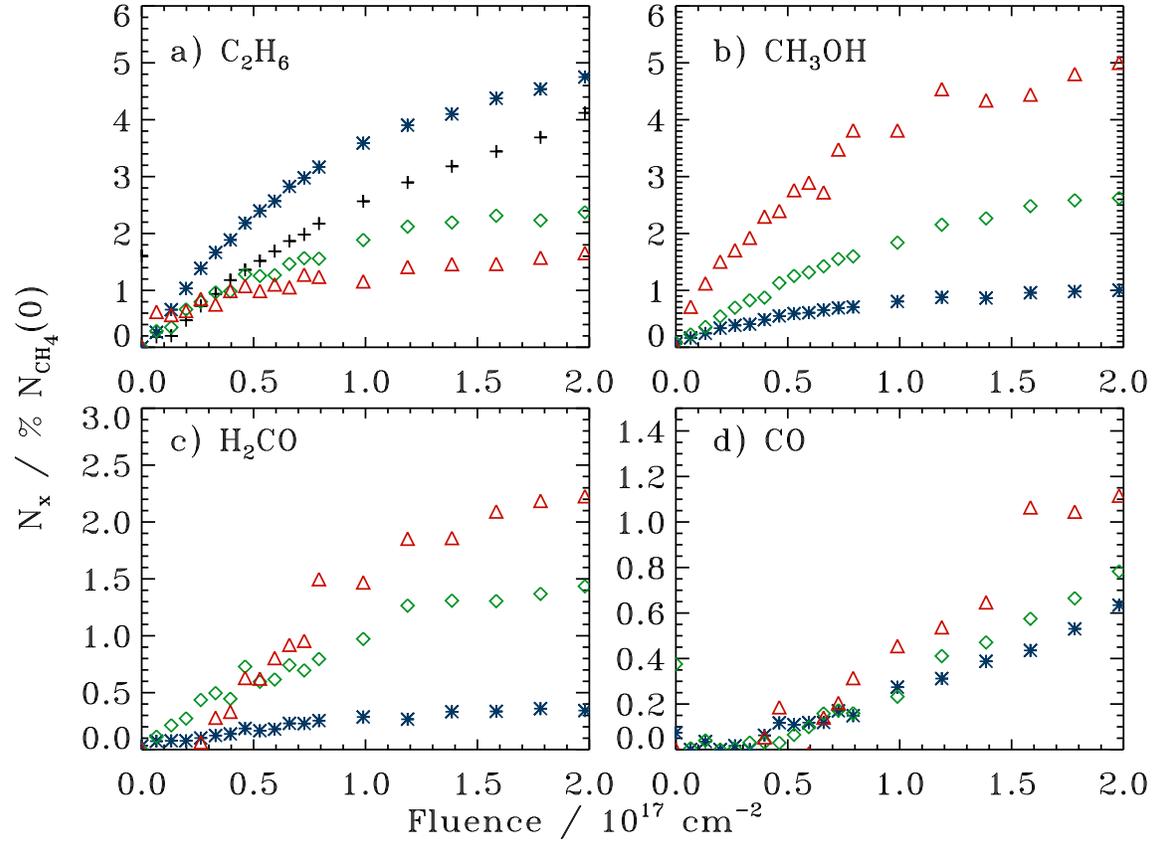}
\caption{The quantified production of C$_2$H$_6$, CH$_3$OH, H$_2$CO and CO in the different mixtures (crosses = pure CH$_4$, blue stars = H$_2$O:CH$_4$ 1:3, green diamonds = 2:1, red triangles = 4:1) as a function of UV fluence. \label{fig:conc_quant}}
\end{figure}

Four of the detected photoproducts, C$_2$H$_6$, CH$_3$OH, H$_2$CO and CO, form abundantly enough to quantify their production as a function of fluence during UV irradiation (Fig. \ref{fig:conc_quant}). C$_2$H$_6$ and CH$_3$OH form immediately upon irradiation in all experiments. The H$_2$CO production is delayed in the H$_2$O-dominated mixture and CO production does not begin before $5\times10^{16}$ photons cm$^{-2}$ in any mixture as expected for a reaction pathway through multiple photodissociation events.

The initial photoproduction rates of C$_2$H$_6$, CH$_3$OH and H$_2$CO are calculated from the abundance growth during the first $4\times10^{16}$ photons cm$^{-2}$ when the growth is still roughly linear. Table \ref{tab:prod_cs} lists the resulting formation 'cross sections'. These numbers do not represent physical photoproduction cross sections, since diffusion is required for their formation, it is however a convenient way of parameterizing their formation. The total formation cross section of complex molecules, in number of product molecules with respect to CH$_4$, increases somewhat with H$_2$O concentration. When taking into account that some products require multiple CHx fragments, an approximately constant fraction of the original CH$_4$ is incorporated into more complex molecules at all concentrations. Furthermore, the CH$_4$ photodestruction cross section increases with an order of magnitude, between pure CH$_4$ and the 4:1 mixture and the formation efficiencies normalized to the photodestruction cross sections thus {\it decrease} significantly with H$_2$O concentration. The presence of H$_2$O then slows down the chemistry of the radicals in the ice, even though this effect is `compensated for' under laboratory time scales by the even more efficient slow-down of back reactions into CH$_4$.

\subsection{Warm up of irradiated ices}

\begin{figure}
\plotone{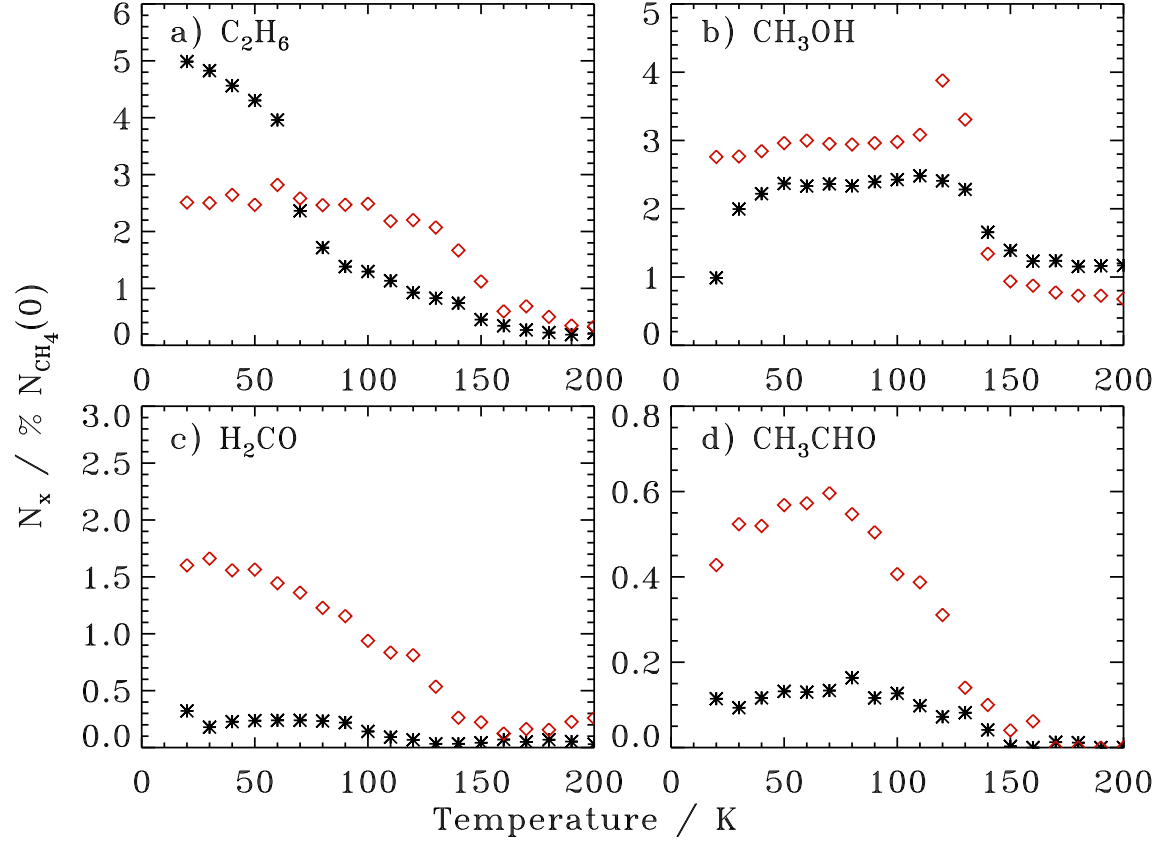}
\caption{The quantified production and destruction of C$_2$H$_6$, CH$_3$OH, H$_2$CO and CH$_3$CHO in two of the mixtures (stars = H$_2$O:CH$_4$ 1:3, diamonds = 2:1) as a function of temperature during warm-up of ices irradiated at 20~K. \label{fig:conc_quant_tpd}}
\end{figure}

During warm-up of the UV-irradiated ices, formation of new species together with sequential desorption change the ice mixture with temperature (Fig. \ref{fig:conc_quant_tpd}). In the H$_2$O-poor 1:3 mixture CH$_3$OH is the only molecule that forms perceptively during warm-up, in the 2:1 mixture CH$_3$CHO is alone to form until reformation of the H$_2$O-ice network around 130~K. The different formation patterns in the H$_2$O-poor and H$_2$O-rich ices suggest different diffusion environments. Specifically, OH diffusion seems hindered in the H$_2$O-rich ice, while HCO can still diffuse and react with CH$_3$ to form CH$_3$CHO. All species are partially trapped in the H$_2$O ice, so that even C$_2$H$_6$ desorption is only complete around the H$_2$O desorption temperature of $\sim$150~K

\subsection{Ice temperature effects during irradiation at 20--60~K}

\begin{figure}
\plotone{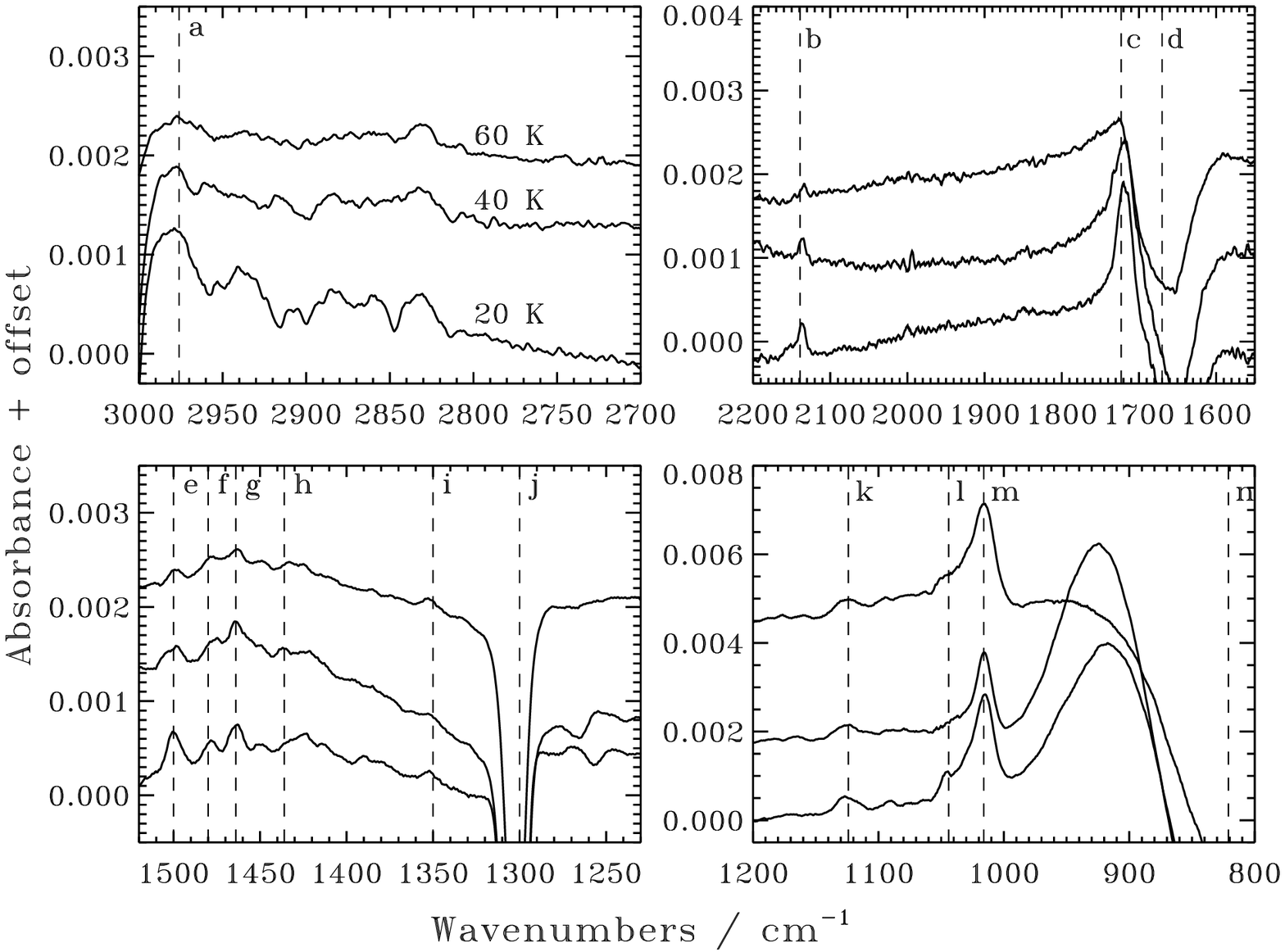}
\caption{Difference spectra of irradiated H$_2$O:CH$_4$ 4-6:1 ice mixtures at 20--60~K after a UV fluence of $2.3\times10^{17}$ photons cm$^{-2}$. The marked peaks are identified with a) C$_2$H$_6$, b) CO, c) XCHO, d) H$_2$O, e) H$_2$CO, f) CH$_3$OH?, g) C$_2$H$_6$, h) C$_2$H$_4$, i) CH$_3$CHO, j) CH$_4$, k) CH$_3$OH, l) CH$_3$CH$_2$OH, m) CH$_3$OH and n) C$_2$H$_4$. \label{fig:temp_spec}}
\end{figure}

\begin{figure}
\plotone{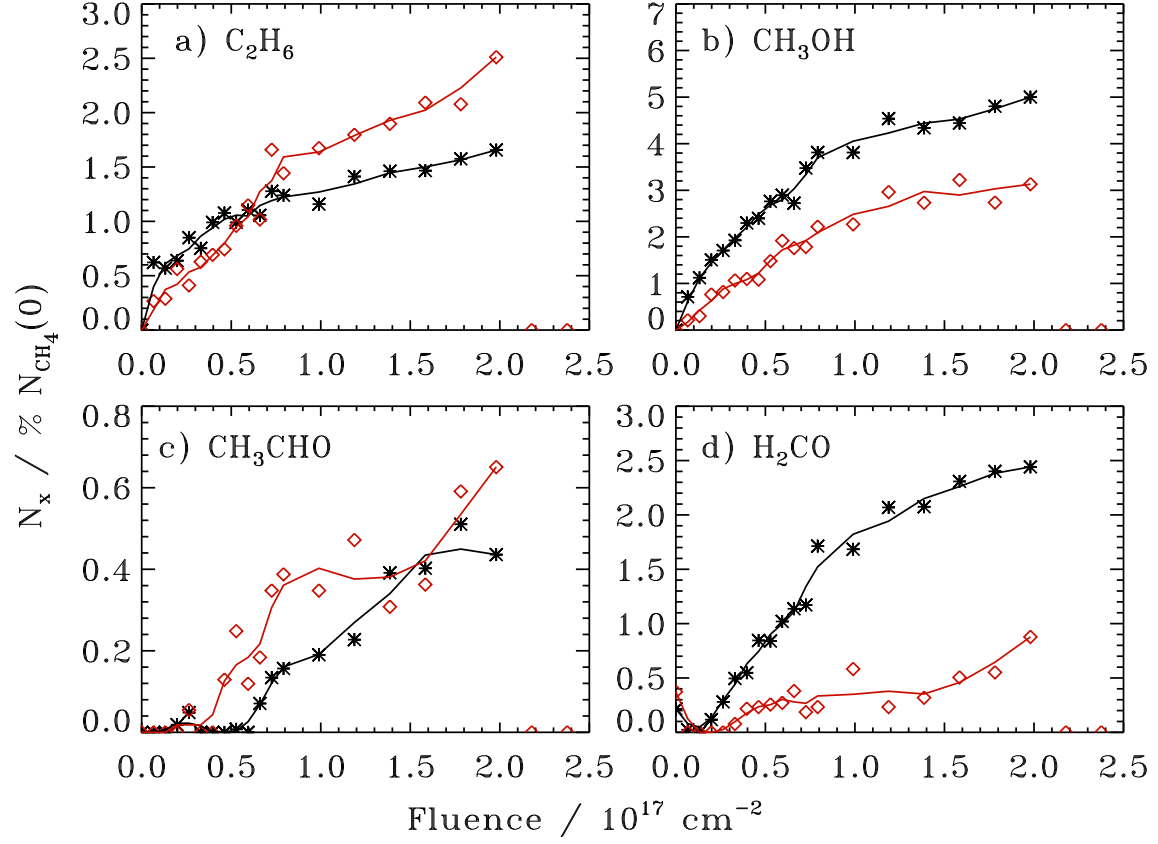}
\caption{The quantified production of C$_2$H$_6$, CH$_3$OH, CH$_3$CHO, H$_2$CO and CH$_3$CH$_2$OH in H$_2$O:CH$_4$  4-6:1 mixtures at 20~K (stars) and 60~K (red diamonds) as a function of UV fluence. The lines shows the 2--3 data point averages. \label{fig:temp_quant}}
\end{figure}

Increasing the ice temperature should speed up the thermal diffusion of radicals in the ice and thus  the ice chemistry. Figure \ref{fig:temp_spec} shows that the most notable changes in the irradiated ice spectra at 20, 40 and 60~K are instead due to outgassing or a decreased production of volatile species, such as H$_2$CO and CO, at the higher temperatures. The decreased production of these secondary and tertiary dissociation products can be understood if the intermediate radicals, e.g. HCO diffuse fast enough at higher temperatures to react with another radical before absorbing a second photon and dissociating to CO. This scenario is supported by an earlier onset in the  CH$_3$CHO production at 60~K compared to 20~K (Fig. \ref{fig:temp_quant}), and also consistent with the warm-up results that reveal HCO diffusion above 30~K. 

In contrast, there is no evidence for an increased formation efficiency of CH$_3$OH and CH$_3$CH$_2$OH or for the formation of more complex molecules such as (CH$_2$OH)$_2$. Rather the CH$_3$OH production cross section decreases from 5.6 to 3.3 $10^{-18}$ cm$^2$ / N$_{\rm CH4}(0)$  between 20 and 60~K. The lack of a positive temperature effect on the CH$_3$OH production, illustrated in Fig.~\ref{fig:temp_quant}, suggests that CH$_3$OH and CH$_3$CH$_2$OH formation depends mainly on CH$_3$ diffusion that is efficient already at 20~K; OH and CH$_2$OH diffusion must be too slow, even at 60~K, in a H$_2$O-dominated ice to affect their formation. In addition the decreasing cross section with temperature is probably due to thermal desorption of some of the produced CH$_3$ at 60~K before it has time to react.

\section{Calculations of binding energies}

To aid the interpretation of the experiments, we calculated the  binding energies of radical-H$_2$O complexes using the CCSD(T) (coupled cluster with single, double, and perturbative triple excitations) method with the aug-cc-pVTZ basis set \citep{Kendall92}. In order to correct for the Basis Set Superposition Error (BSSE) the Boys-Bernardi counterpoise method \citep{Boys70} was used. The internal geometries of the interacting species were kept constant while a large number of different relative orientations and intermolecular distances were scanned. The thus located lowest energy minima were used to calculate the binding energies. The reported energies are electronic energies relative to the infinitely separated interaction partners. No account of zero-point energy was made. All calculations were performed using the Gaussian 03 program package \citep{Frisch03}.

The calculations were carried out for the predicted three most common radicals in the ice, OH, HCO and CH$_3$. The resulting binding energies are 0.23 eV (2700~K) for H$_2$O-OH, 0.11 eV (1300~K) for H$_2$O:HCO and 0.07 eV (800~K) for H$_2$O-CH$_3$. The H$_2$O-OH result is in good agreement with other recent high-level calculations \citep{Soloveichik10}

\section{Discussion}

\subsection{The influence of H$_2$O}

Under laboratory conditions the most important effect of increasing the H$_2$O concentration is to drive the chemistry away from C$_x$H$_y$ chains and toward products with a higher and higher concentration of OH. The same effect was noted by \citet{Moore98} during proton bombardment of H$_2$O:CH$_4$ ice mixtures. This can be understood from simple rate equations where the production rates of various molecules are proportional to the production rates of CH$_3$/CH$_2$ and OH in the ice, i.e. the photodestruction cross sections of CH$_4$ and H$_2$O. The absolute values of these reaction rates also depend on the diffusion rates of radicals. If the destruction and diffusion rates had been constant, the relations shown in Fig. \ref{fig:conc_fin} could have been used to derive the relative diffusion barriers directly. 

H$_2$O does, however,  change the CH$_4$ photodestruction cross section. Since this increased destruction efficiency is not accompanied by an equally large increase in production efficiencies of new molecules, H$_2$O must slow down the overall diffusion of radicals in the ice to the point of trapping them. H is not expected to be trapped at 20~K, however, and the existence of larger radicals in the ice indicates that many of the H atoms formed during dissociation either escapes from the ice or react to form H$_2$. A significant amount of the produced heavier radicals, especially CH$_3$, must also remain mobile in or on top of the ice to account for the production of any complex molecules during irradiation at 20~K in the H$_2$O-rich ices. Without detailed modeling it is not possible to estimate the relative importances of surface and bulk diffusion for the chemistry. It is however unlikely that all radicals are trapped in the bulk of the ice at 20~K, since even in the most dilute mixtures $\sim$20\% of the dissociated CH$_4$ molecules recombine to form more complex molecules. This corresponds to at least 8~ML of chemically active ice in a 40~ML thick ice and the ice would thus have to be very porous to explain all chemistry with diffusion on external and internal (in large pores) ice surfaces. It would also be strange if diffusion in H$_2$O ices is fundamentally different compared to diffusion in CH$_3$OH ices, where bulk diffusion is definitely required to explain the ice photochemistry results \citep{Oberg09d}, considering the similar behavior of the two ices during segregation studies \citep{Ehrenfreund99,Bernstein05}. 

Trapping of some of the radicals in the ice bulk is however needed to explain that up to 80\% of the dissociated CH$_4$ is not converted into more complex molecules in the H$_2$O:CH$_4$ 5:1 ices. This is consistent with the commonly observed trapping of molecules such as CO and CO$_2$ in H$_2$O-rich ices \citep{Collings04}. It may also explain some earlier results; trapping of CH$_3$OH photofragments in the H$_2$O:CH$_3$OH 10:1 ice mixture in \citet{Krim09} would prevent the radicals reacting with each other and thus favor dissociation into smaller and smaller fragments (CH$_2$OH, H$_2$CO, HCO, CO), explaining their result of high H$_2$CO and CO yields in H$_2$O-rich ices without invoking efficient H abstraction by OH radicals. %Quantitative modeling and more targeted experiments, focused on ice thickness and compactness effects, are however required to confirm the importance of trapping during H$_2$O-ice chemistry.

When the ice is heated some of the bound radicals overcome their entrapment and react to form new species. In all H$_2$O:CH$_4$ ice mixtures the most abundant radicals are expected to be CH$_3$ (see below) and OH, the two major photoproducts of CH$_4$ and H$_2$O.  CH$_3$OH is thus always expected to form more abundantly during warm-up of the irradiated ices than e.g. CH$_3$CHO, consistent with the results in Fig. \ref{fig:conc_quant} for the H$_2$O:CH$_4$ 1:3 experiment. In the 2:1 mixture almost no CH$_3$OH forms initially during warm-up, however. Instead a significant amount of CH$_3$CHO forms, followed later by CH$_3$OH formation during the H$_2$O ice reformation temperature of $\sim$120--130~K. An increasing H$_2$O concentration thus increases the OH diffusion barrier relative to the HCO diffusion barrier. This is physically reasonable since OH and H$_2$O are expected to bond stronger than HCO and H$_2$O. The importance of the relative diffusion barriers of OH and HCO to drive the chemistry demonstrates that radical-radical reactions in ice is only efficient when both radicals (i.e. CH$_3$+OH or CH$_3$+HCO) taking part in the reaction are mobile. 

\subsection{Toward quantifying the H$_2$O:CH$_4$ photochemistry}

CH$_4$ is known to dissociate into a number of different fragments during UV photolysis in the gas phase. The dissociation cross section is wavelength dependent, e.g. almost no CH$_2$ is observed to form from dissociation with Ly-$\alpha$ photons \citep{Mordaunt93} even though it is expected to be an important channel at lower photon energies \citep{Slanger82}. By comparing the initial C$_2$H$_6$ and C$_2$H$_4$ formation efficiencies in pure UV-irradiated CH$_4$ ice, it is possible to constrain the CH$_4$ photodissociation branching ratio in ices irradiated with a hydrogen lamp output consisting of both Ly-$\alpha$ photons and broad-band emission.

In the pure CH$_4$ ice the initial growth of both C$_2$H$_6$ and C$_2$H$_4$ is assumed to be caused by CH$_3$+CH$_3$ and CH$_2$+CH$_2$ radical reactions, ignoring second generation photodissociation products. Further assuming that the diffusion barriers of CH$_3$ and CH$_2$ are comparable, the CH$_4$ branching ratio can be constrained from the initial C$_2$H$_6$ and C$_2$H$_4$ production rate, which is $\sim$9:1; \citet{Gerakines96} find a comparable C$_2$H$_6$/C$_2$H$_4$ product ratio (8:1). Since each reaction requires two radicals, the dissociation branching ratio is inferred to be CH$_3$:CH$_2$ $\gtrsim$3:1. This is consistent with a chemistry driven largely by Ly-$\alpha$ photons.

The diffusion barriers are more difficult to constrain directly from the experiment.  The calculated bond strengths of 800, 1300 and 2700~K for the H$_2$O-CH$_3$, --HCO and -OH complexes are not identical to ice diffusion barriers, but the relative values should agree with the experimental results. Indeed, the data show that OH has a higher diffusion barrier than HCO in H$_2$O-dominated ices (efficient thermal diffusion at 30 and 120~K respectively) and that CH$_3$ must be diffusing already at 20~K to explain any complex molecule production at this temperature. The calculated bond strengths are thus a good starting point for models, where the quantified formation rates of molecules can be used to extract diffusion barriers, similarly to what is currently being done for the CH$_3$OH photochemistry (Garrod \& \"Oberg in prep.). Already both calculations and experiments suggest that  the temperature window between diffusion and desorption in H$_2$O-dominated ices is quite small for the investigated radicals, which may explain the low production yield of complex organics during warm-up (a few \% of all dissociated CH$_4$) observed here compared to the efficient production of complex molecules observed during warm-up of irradiated CH$_3$OH-ices \citep{Oberg09d}.

The calculations also show that predicting radical diffusion barriers from the bond strengths of the parent molecule to H$_2$O can be quite reasonable \citep{Garrod08}; the H$_2$O-H$_2$O, H$_2$O-H$_2$CO and H$_2$O-CH$_4$ bond strengths are 0.2, 0.07-0.13 and 0.03 eV, respectively \citep[e.g.][]{Szczesniak93, DelBene73}, which is very similar to the bond strengths calculated for the H$_2$O-radical complex interactions. There thus seems to be a direct dependence between the H$_2$O-molecule and H$_2$O-radical  interactions.

\subsection{Astrophysical implications}

As shown previously for pure and mixed CH$_3$OH ices, UV irradiation of H$_2$O:CH$_4$ ices readily results in the production of complex molecules and this production can be quantified. After a UV fluence corresponding to $\sim10^6$ years in a cloud core, up to 50\% of the original CH$_4$ and 25\% of the NH$_3$ ice have dissociated into radicals that can react into more complex species. At low temperatures most of these radicals are trapped in the ice, i.e. only $\sim$10\% of the original CH$_4$ ice can be converted into complex molecules during UV irradiation at low temperatures. As the protostar turns on and heats the ice many of the trapped radicals will become mobile resulting in a second equally important formation step of complex molecules -- this step will probably be more important in space than in the lab because of the lower heating rate in astrophysical environments. Taking a typical CH$_4$ ice abundance of $\sim$4\% with respect to H$_2$O ice and a H$_2$O ice abundance of $\sim$10$^{-4}$ with respect to H$_2$ toward protostars \citep{Oberg08}, this corresponds to a potential production of complex molecules of $0.04\times10^{-4}\times0.1=4\times10^{-7}$ with respect to H$_2$. This is high enough to contribute significantly to typical hot core abundances of complex molecules $\sim10^{-9}-10^{-6}$ \citep{Bisschop07}.

Several of the CH$_3$OH and H$_2$O:CH$_4$ photochemistry products are the same, e.g. CH$_3$CH$_2$OH and CH$_3$CHO, and the relative importance of the two production pathways during star formation will depend on the relative abundances of CH$_3$OH and CH$_4$ ice on the grains as well as what other ice constituents are mixed with H$_2$O. Both the H$_2$O- and the CO/CH$_3$OH-rich ice phases must clearly be modeled to accurately predict the total formation rates of C- and O-containing complex molecules in space. 

In addition it is crucial to understand the effects of H$_2$O quantitatively to constrain when and where N-containing complex molecules form -- their main formation pathway is probably from NH$_3$ in a H$_2$O-rich ice. While the production of radicals seems to be the most important effect of H$_2$O on the ice photochemistry under laboratory conditions, this is not necessarily the case for the time scales and ice compositions present at star and planet formation. Rather detailed modeling is required that extracts the microscopic properties of the H$_2$O-rich ice chemistry from the laboratory experiments, such as the composition-dependent diffusion barriers, and applies these results to astrochemical models. Only when a simple system, such as H$_2$O:CH$_4$, is understood at this detailed level can we expect to accurately model the more elusive nitrogen chemistry in H$_2$O-rich environments and thus provide predictions of e.g. the prebiotic amino acid production. 

Estimating ice life times is easier than predicting the chemical evolution, though the composition dependence of CH$_4$ photodestruction introduces some uncertainty. Table \ref{tab:ch4_life} list the photodestruction time scales for 20-30~K CH$_4$ ice in pure, 2:1 and 4:1 H$_2$O:CH$_4$ ices subject to a weak UV field induced by cosmic rays, the interstellar radiation field and the 1000 times higher UV flux inferred toward the L1527 outflow \citep{Spaans95}. Outside of protected cloud cores the CH$_4$ ice life time is thus  short compared to most other time scales, especially in H$_2$O-rich ice mixtures relevant for protostars. For large quantities of CH$_4$ ice to survive to thermally desorb around protostars and drive warm-carbon-chain-chemistry as suggested by \citep{Sakai08} then requires that the ices are well protected up until that point. This may be a reason for the low number of observations of such a chemistry during the protostellar stage.

\section{Conclusions}

\begin{enumerate}
\item The photodestruction cross section of NH$_3$ and CH$_4$ in H$_2$O ice mixtures increases with H$_2$O concentration by up to a factor of 10 with implications for the CH$_4$ ice lifetime in protostellar envelopes. Simultaneously the production efficiency of stable molecules in the H$_2$O:CH$_4$ ice increases by a factor of two. This is explained by  trapping of radicals in the H$_2$O-ice matrix, which prevents back reactions between e.g. CH$_3$ and H. 
\item The H$_2$O:CH$_4$ photochemistry produces C$_2$H$_6$, CH$_3$OH and CH$_3$CH$_2$OH as a function of CH$_4$ concentration, in a way that can be directly related to the number of OH and CH$_2$/CH$_3$ groups in each product. The production of CH$_3$CHO, H$_2$CO and CO has a fluence delay consistent with formation through more steps compared to e.g. CH$_3$OH.
\item On laboratory time scales increasing the ice temperature from 20 to 60~K has a limited effect on the chemistry, probably because HCO is the only important reactant that becomes mobile in this temperature interval in H$_2$O-dominated ice mixtures.
\item The CH$_4$ photodissociation branching ratio into CH$_3$ and CH$_2$ is $\gtrsim$3:1, as derived from the relative production efficiencies of C$_2$H$_6$ and C$_2$H$_4$.
\item OH diffusion is fast in the  H$_2$O:CH$_4$ 1:3 mixture, but not present in the H$_2$O-dominated ice mixtures below 100~K. HCO diffusion is possible above 30~K in the H$_2$O-rich ice mixtures and CH$_3$ diffusion seems efficient at 20~K in all mixtures, consistent with  calculated H$_2$O-radical complex bond strengths. OH diffusion is thus more severely slowed down with H$_2$O concentration than the other radicals, which counteracts the higher OH production rate in H$_2$O-richer ices. 
\item Because H$_2$O is both a source of OH radicals and preferentially trap OH compared to more volatile radicals, predicting the photoproduction branching ratio of different complex molecules in space requires microscopic modeling of the H$_2$O:CH$_4$ experiments rather than direct comparison with experiments. 
\end{enumerate}

\acknowledgments

This work has benefitted from discussions with Herma Cuppen and Robin Garrod. Support for KIO is provided by NASA through Hubble Fellowship grant  awarded by the Space Telescope Science Institute, which is operated by the Association
of Universities for Research in Astronomy, Inc., for NASA. Astrochemistry in Leiden is supported by a SPINOZA grant of the Netherlands Organization for Scientific Research (NWO) and NOVA. The work by SA was partly funded by a NWO-CW Top grant. Some of the electronic structure calculations were performed at C3SE (Chalmers Centre for Computational Science and Engineering) computing resources.

\newpage

\newpage

\newpage

\newpage

\newpage

\newpage

\newpage

\newpage

\newpage

\newpage

\newpage

\begin{deluxetable}{lcccc}
%\rotate
\tablecolumns{5} 
\tabletypesize{\scriptsize}
\tablecaption{The photochemistry experiments.\label{tab:n-exps}}
\tablewidth{0pt}
\tablehead{
\colhead{Ice} & \colhead{H$_2$O:x} & \colhead{Temp. (K)} & \colhead{Thick. (ML)} & \colhead{cross section (10$^{-18}$ cm$^2$)} 
}
\startdata
CH$_4$			&0:1	&30	&47		&0.5[0.3]\\%040809
H$_2$O:CH$_4$	&1:3	&20	&42		&2.8[1.4]\\%050609
H$_2$O:CH$_4$	&2:1	&20	&37		&2.9[1.5]\\%061609
H$_2$O:CH$_4$	&4:1&20	&37(29:8)	&4.9[2.5]\\%072909
H$_2$O:CH$_4$	&4:1&40	&38(30:8)	&4.6[2.3]\\%081009
H$_2$O:CH$_4$	&5:1&60	&38(32:6)	&3.9[2.0]\\%08
NH$_3$			&0:1	&30	&51		&1.4[0.7]\\%050909
H$_2$O:NH$_3$	&1:1	&20	&54		&1.5[0.8]\\%
H$_2$O:NH$_3$	&4:1	&20	&43		&5.0[2.5]\\%061409
CO$_2$			&0:1	&30	&15\\%042809
H$_2$O:CO$_2$	&6:1	&20	&35\\%061709
\enddata
\end{deluxetable}

\begin{table}[ht]
\begin{center}
{\small
\caption{Ice infrared spectral features used for quantification.\label{tab:bands}}
\begin{tabular}{lccc}
\hline \hline
Species & Band (cm$^{-1}$) & Band strength$^{\rm a}$ (cm$^{-1}$) & Reference\\
\hline
CH$_4$			&1300	&6.1$\times10^{-18}$	&\citet{Moore98}\\
NH$_3$			&1070	&1.7$\times10^{-17}$	&\citet{dHendecourt86}\\
CO$_2$			&2343	&7.6$\times10^{-17}$	&\citet{Gerakines95}\\
H$_2$O			&1670	&1.2$\times10^{-17}$	&\citet{Gerakines95}\\
C$_2$H$_6$		&2976	&1.1$\times10^{-17}$	&\citet{Moore98}$^{\rm b}$\\
				&821	&1.9$\times10^{-18}$	&\citet{Pearl91}\\
C$_2$H$_4$		&1436	&2.9$\times10^{-18}$	&\citet{Moore98}$^{\rm b}$\\
H$_2$CO			&1500	&3.9$\times10^{-18}$	&\citet{Schutte93}\\
CH$_3$OH		&1026	&2.8$\times10^{-17}$	&\citet{dHendecourt86}\\
CH$_3$CH$_2$OH	&1044	&7.3$\times10^{-18}$	&\citet{Moore98}$^{\rm b}$\\
CH$_3$CHO		&1350	&6.1$\times10^{-18}$	&\citet{Moore98}$^{\rm b}$\\
CO				&2139	&1.1$\times10^{-17}$	&\citet{Gerakines95}\\
\hline
\end{tabular}
\\$^{\rm a}$ The band strengths are known within $\sim$20--30\% when comparing results from different references, ice mixtures and ice temperatures.\\ 
$^{\rm b}$ In a H$_2$O ice matrix. 
}
\end{center}
\end{table}

\begin{table}[ht]
\begin{center}
{\small
\caption{Normalized product 'cross sections' ($10^{-18}$ cm$^2$ / N$_{\rm CH4}(0)$) at 20 K for H$_2$O:CH$_4$ ices. Fit uncertainties are in brackets. \label{tab:prod_cs}}
\begin{tabular}{lcccc}
\hline \hline
Product & Pure CH$_4$ &1:3 & 2:1 & 4:1\\
\hline
C$_2$H$_6$	&0.30[0.10]	&0.52[0.18]	&0.29[0.10]	&$\lesssim$0.20\\
C$_2$H$_4$	&0.04[0.01]\\
CH$_3$OH	&--		&0.11[0.04]	&0.23[0.08]	&0.56[0.19]\\
H$_2$CO		&--		&$<$0.02		&0.15[0.05]	&0.10[0.04]\\
\tableline
Total$^{\rm a}$ 			&0.34	&0.63		&0.67		&$\lesssim$0.86\\
Total / \%$^{\rm b}$ 	&$\sim$100		&41			&33			&$\lesssim$22\\
\hline
\end{tabular}
\\$^{\rm a}$ The total formation cross section of new molecules. \\ 
$^{\rm b}$ \% of the photodestruction cross section that is accounted for by the total formation cross section taking into account that some molecules require multiple CH$_x$ fragments to form. \\ 
}
\end{center}
\end{table}

\begin{table}[ht]
\begin{center}
{\small
\caption{CH$_4$ ice life times in years. \label{tab:ch4_life}}
\begin{tabular}{lccc}
\hline \hline
UV flux (cm$^{-2}$ s$^{-1}$) & pure CH$_4$ & 2:1 & 4:1\\
\hline
10$^4$ (cloud core)		&$6\times10^6$	&$1\times10^6$&$6\times10^5$\\
10$^8$ (cloud edge)	&$6\times10^2$	&$1\times10^2$&$6\times10^1$\\
10$^{11}$ (outflow cavity)	&0.6				&0.1			&0.06\\
\hline
\end{tabular}
}
\end{center}
\end{table}


\begin{thebibliography}{44}

\bibitem[{{Belloche} {et~al.}(2009){Belloche}, {Garrod}, {M{\"u}ller},
  {Menten}, {Comito}, \& {Schilke}}]{Belloche09}
{Belloche}, A., {Garrod}, R.~T., {M{\"u}ller}, H.~S.~P., {et~al.} 2009, \aap,
  499, 215

\bibitem[{Bene(1973)}]{DelBene73}
Bene, J. E.~D. 1973, Chemical Physics Letters, 23, 287

\bibitem[{{Bergin} {et~al.}(2002){Bergin}, {Alves}, {Huard}, \&
  {Lada}}]{Bergin02}
{Bergin}, E.~A., {Alves}, J., {Huard}, T., \& {Lada}, C.~J. 2002, \apjl, 570,
  L101

\bibitem[{{Bernstein} {et~al.}(2005){Bernstein}, {Cruikshank}, \&
  {Sandford}}]{Bernstein05}
{Bernstein}, M.~P., {Cruikshank}, D.~P., \& {Sandford}, S.~A. 2005, Icarus,
  179, 527

\bibitem[{{Bisschop} {et~al.}(2007){Bisschop}, {J{\o}rgensen}, {van Dishoeck},
  \& {de Wachter}}]{Bisschop07}
{Bisschop}, S.~E., {J{\o}rgensen}, J.~K., {van Dishoeck}, E.~F., \& {de
  Wachter}, E.~B.~M. 2007, \aap, 465, 913

\bibitem[{{Boogert} {et~al.}(2008){Boogert}, {Pontoppidan}, {Knez}, {Lahuis},
  {Kessler-Silacci}, {van Dishoeck}, {Blake}, {Augereau}, {Bisschop},
  {Bottinelli}, {Brooke}, {Brown}, {Crapsi}, {Evans}, {Fraser}, {Geers},
  {Huard}, {J{\o}rgensen}, {{\"O}berg}, {Allen}, {Harvey}, {Koerner}, {Mundy},
  {Padgett}, {Sargent}, \& {Stapelfeldt}}]{Boogert08}
{Boogert}, A.~C.~A., {Pontoppidan}, K.~M., {Knez}, C., {et~al.} 2008, \apj,
  678, 985

\bibitem[{{Boys} \& {Bernardi}(1970)}]{Boys70}
{Boys}, S.~F. \& {Bernardi}, F. 1970, Mol. Phy., 19, 553

\bibitem[{{Collings} {et~al.}(2004){Collings}, {Anderson}, {Chen}, {Dever},
  {Viti}, {Williams}, \& {McCoustra}}]{Collings04}
{Collings}, M.~P., {Anderson}, M.~A., {Chen}, R., {et~al.} 2004, \mnras, 354,
  1133

\bibitem[{{Collings} {et~al.}(2003){Collings}, {Dever}, {Fraser}, \&
  {McCoustra}}]{Collings03}
{Collings}, M.~P., {Dever}, J.~W., {Fraser}, H.~J., \& {McCoustra}, M.~R.~S.
  2003, \apss, 285, 633

\bibitem[{{Cottin} {et~al.}(2003){Cottin}, {Moore}, \&
  {B{\'e}nilan}}]{Cottin03}
{Cottin}, H., {Moore}, M.~H., \& {B{\'e}nilan}, Y. 2003, \apj, 590, 874

\bibitem[{{Crovisier} {et~al.}(2004){Crovisier}, {Bockel{\'e}e-Morvan},
  {Biver}, {Colom}, {Despois}, \& {Lis}}]{Crovisier04}
{Crovisier}, J., {Bockel{\'e}e-Morvan}, D., {Biver}, N., {et~al.} 2004, \aap,
  418, L35

\bibitem[{{D'Hendecourt} \& {Allamandola}(1986)}]{dHendecourt86}
{D'Hendecourt}, L.~B. \& {Allamandola}, L.~J. 1986, \aaps, 64, 453

\bibitem[{{Ehrenfreund} {et~al.}(1999){Ehrenfreund}, {Kerkhof}, {Schutte},
  {Boogert}, {Gerakines}, {Dartois}, {D'Hendecourt}, {Tielens}, {van Dishoeck},
  \& {Whittet}}]{Ehrenfreund99}
{Ehrenfreund}, P., {Kerkhof}, O., {Schutte}, W.~A., {et~al.} 1999, \aap, 350,
  240

\bibitem[{{Frisch} {et~al.}(2004)}]{Frisch03}
{Frisch}, M.~J. {et~al.} 2004, {Gaussian 03 program package (Rev. E.01; Wallingford:
  Gaussian, Inc.)}

\bibitem[{{Garrod} {et~al.}(2008){Garrod}, {Weaver}, \& {Herbst}}]{Garrod08}
{Garrod}, R.~T., {Weaver}, S.~L.~W., \& {Herbst}, E. 2008, \apj, 682, 283

\bibitem[{{Gerakines} {et~al.}(1996){Gerakines}, {Schutte}, \&
  {Ehrenfreund}}]{Gerakines96}
{Gerakines}, P.~A., {Schutte}, W.~A., \& {Ehrenfreund}, P. 1996, \aap, 312, 289

\bibitem[{{Gerakines} {et~al.}(1995){Gerakines}, {Schutte}, {Greenberg}, \&
  {van Dishoeck}}]{Gerakines95}
{Gerakines}, P.~A., {Schutte}, W.~A., {Greenberg}, J.~M., \& {van Dishoeck},
  E.~F. 1995, \aap, 296, 810

\bibitem[{{Greenberg} \& {Hong}(1974)}]{Greenberg74}
{Greenberg}, J.~M. \& {Hong}, S.-S. 1974, in IAU Symposium, Vol.~60, Galactic
  Radio Astronomy, ed. F.~J. {Kerr} \& S.~C. {Simonson}, 155--177

\bibitem[{{Hagen} {et~al.}(1979){Hagen}, {Allamandola}, \&
  {Greenberg}}]{Hagen79}
{Hagen}, W., {Allamandola}, L.~J., \& {Greenberg}, J.~M. 1979, \apss, 65, 215

\bibitem[{{Ioppolo} {et~al.}(2008){Ioppolo}, {Cuppen}, {Romanzin}, {van
  Dishoeck}, \& {Linnartz}}]{Ioppolo08}
{Ioppolo}, S., {Cuppen}, H.~M., {Romanzin}, C., {van Dishoeck}, E.~F., \&
  {Linnartz}, H. 2008, \apj, 686, 1474

\bibitem[{Kendall {et~al.}(1992)Kendall, Thom H.~Dunning, \&
  Harrison}]{Kendall92}
Kendall, R.~A., Thom H.~Dunning, J., \& Harrison, R.~J. 1992, The Journal of
  Chemical Physics, 96, 6796

\bibitem[{{Krim} {et~al.}(2009){Krim}, {Lasne}, {Laffon}, \& {Parent}}]{Krim09}
{Krim}, L., {Lasne}, J., {Laffon}, C., \& {Parent}, P. 2009, The Journal of
  Physical Chemistry A, 113, 8979

\bibitem[{{Moore} \& {Hudson}(1998)}]{Moore98}
{Moore}, M.~H. \& {Hudson}, R.~L. 1998, Icarus, 135, 518

\bibitem[{Mordaunt {et~al.}(1993)Mordaunt, Lambert, Morley, Ashfold, Dixon,
  Western, Schnieder, \& Welge}]{Mordaunt93}
Mordaunt, D.~H., Lambert, I.~R., Morley, G.~P., {et~al.} 1993, J. Chem. Ph.,
  98, 2054

\bibitem[{{Mu{\~n}oz Caro} \& {Schutte}(2003)}]{MunozCaro03}
{Mu{\~n}oz Caro}, G.~M. \& {Schutte}, W.~A. 2003, \aap, 412, 121

\bibitem[{{{\"O}berg} {et~al.}(2008){{\"O}berg}, {Boogert}, {Pontoppidan},
  {Blake}, {Evans}, {Lahuis}, \& {van Dishoeck}}]{Oberg08}
{{\"O}berg}, K.~I., {Boogert}, A.~C.~A., {Pontoppidan}, K.~M., {et~al.} 2008,
  \apj, 678, 1032

\bibitem[{{{\"O}berg} {et~al.}(2007){{\"O}berg}, {Fraser}, {Boogert},
  {Bisschop}, {Fuchs}, {van Dishoeck}, \& {Linnartz}}]{Oberg07a}
{{\"O}berg}, K.~I., {Fraser}, H.~J., {Boogert}, A.~C.~A., {et~al.} 2007, \aap,
  462, 1187

\bibitem[{{{\"O}berg} {et~al.}(2009{\natexlab{a}}){{\"O}berg}, {Garrod}, {van
  Dishoeck}, \& {Linnartz}}]{Oberg09d}
{{\"O}berg}, K.~I., {Garrod}, R.~T., {van Dishoeck}, E.~F., \& {Linnartz}, H.
  2009{\natexlab{a}}, \aap, 504, 891

\bibitem[{{{\"O}berg} {et~al.}(2009{\natexlab{b}}){{\"O}berg}, {Linnartz},
  {Visser}, \& {van Dishoeck}}]{Oberg09c}
{{\"O}berg}, K.~I., {Linnartz}, H., {Visser}, R., \& {van Dishoeck}, E.~F.
  2009{\natexlab{b}}, \apj, 693, 1209

\bibitem[{{{\"O}berg} {et~al.}(2005){{\"O}berg}, {van Broekhuizen}, {Fraser},
  {Bisschop}, {van Dishoeck}, \& {Schlemmer}}]{Oberg05}
{{\"O}berg}, K.~I., {van Broekhuizen}, F., {Fraser}, H.~J., {et~al.} 2005,
  \apjl, 621, L33

\bibitem[{{{\"O}berg} {et~al.}(2009{\natexlab{c}}){{\"O}berg}, {van Dishoeck},
  \& {Linnartz}}]{Oberg09b}
{{\"O}berg}, K.~I., {van Dishoeck}, E.~F., \& {Linnartz}, H.
  2009{\natexlab{c}}, \aap, 496, 281

\bibitem[{{Pearl} {et~al.}(1991){Pearl}, {Ngoh}, {Ospina}, \&
  {Khanna}}]{Pearl91}
{Pearl}, J., {Ngoh}, M., {Ospina}, M., \& {Khanna}, R. 1991, \jgr, 96, 17477

\bibitem[{{Pontoppidan}(2006)}]{Pontoppidan06}
{Pontoppidan}, K.~M. 2006, \aap, 453, L47

\bibitem[{{Pontoppidan} {et~al.}(2003){Pontoppidan}, {Fraser}, {Dartois},
  {Thi}, {van Dishoeck}, {Boogert}, {d'Hendecourt}, {Tielens}, \&
  {Bisschop}}]{Pontoppidan03}
{Pontoppidan}, K.~M., {Fraser}, H.~J., {Dartois}, E., {et~al.} 2003, \aap, 408,
  981

\bibitem[{{Sakai} {et~al.}(2008){Sakai}, {Sakai}, {Hirota}, \&
  {Yamamoto}}]{Sakai08}
{Sakai}, N., {Sakai}, T., {Hirota}, T., \& {Yamamoto}, S. 2008, \apj, 672, 371

\bibitem[{{Schutte} {et~al.}(1993){Schutte}, {Allamandola}, \&
  {Sandford}}]{Schutte93}
{Schutte}, W.~A., {Allamandola}, L.~J., \& {Sandford}, S.~A. 1993, Icarus, 104,
  118

\bibitem[{{Shen} {et~al.}(2004){Shen}, {Greenberg}, {Schutte}, \& {van
  Dishoeck}}]{Shen04}
{Shen}, C.~J., {Greenberg}, J.~M., {Schutte}, W.~A., \& {van Dishoeck}, E.~F.
  2004, \aap, 415, 203

\bibitem[{Slanger \& Black(1982)}]{Slanger82}
Slanger, T.~G. \& Black, G. 1982, \jcp, 77, 2432

\bibitem[{Soloveichik {et~al.}(2010)Soloveichik, O'Donnell, Lester, Francisco,
  \& McCoy}]{Soloveichik10}
Soloveichik, P., O'Donnell, B.~A., Lester, M.~I., Francisco, J.~S., \& McCoy,
  A.~B. 2010, J. Ph. Chem. A, 114, 1529

\bibitem[{{Spaans} {et~al.}(1995){Spaans}, {Hogerheijde}, {Mundy}, \& {van
  Dishoeck}}]{Spaans95}
{Spaans}, M., {Hogerheijde}, M.~R., {Mundy}, L.~G., \& {van Dishoeck}, E.~F.
  1995, \apjl, 455, L167+

\bibitem[{Szczesniak {et~al.}(1993)Szczesniak, Chalasinski, Cybulski, \&
  Cieplak}]{Szczesniak93}
Szczesniak, M.~M., Chalasinski, G., Cybulski, S.~M., \& Cieplak, P. 1993, The
  Journal of Chemical Physics, 98, 3078

\bibitem[{{Tielens} \& {Hagen}(1982)}]{Tielens82}
{Tielens}, A.~G.~G.~M. \& {Hagen}, W. 1982, \aap, 114, 245

\bibitem[{{van Dishoeck}(1988)}]{vanDishoeck88}
{van Dishoeck}, E.~F. 1988, in Rate Coefficients in Astrochemistry., ed. T.~J.
  {Millar} \& D.~A. {Williams}, 49

\bibitem[{{Watanabe} \& {Kouchi}(2002)}]{Watanabe02}
{Watanabe}, N. \& {Kouchi}, A. 2002, \apj, 567, 651

\end{thebibliography}
\end{document}